\documentclass[%
 reprint,
superscriptaddress,
 amsmath,amssymb,
prl,
floatfix,
]{revtex4-2}

\usepackage{graphicx}
\usepackage{dcolumn}
\usepackage{bm}

\usepackage[dvipsnames]{xcolor}

\usepackage[colorlinks={true}, citecolor={BlueViolet}, filecolor={BlueViolet}, linkcolor={BlueViolet}, urlcolor={BlueViolet}]{hyperref}
\usepackage[caption=false]{subfig}

\newcommand{\ketbra}[2]{\left| #1 \right\rangle\!\left\langle #2 \right|}

\usepackage[utf8]{inputenc}
\usepackage[normalem]{ulem}
\usepackage{amsmath}
\usepackage{mathrsfs} 
\begin{document}

\preprint{APS/123-QED}

\title{Pushing the Boundaries: Interferometric Mass Photometry at the Quantum Limit of Sensitivity} 
\author{Fabian M\"uller}
\affiliation{Institut f\"{u}r Theoretische Physik, Eberhard Karls Universit\"{a}t T\"{u}bingen, 72076 T\"{u}bingen, Germany}

\author{Emre K\"ose}
\affiliation{Institut f\"{u}r Theoretische Physik, Eberhard Karls Universit\"{a}t T\"{u}bingen, 72076 T\"{u}bingen, Germany}

\author{Alfred J. Meixner}
\affiliation{Institut f\"{u}r Physikalische und Theoretische Chemie, Eberhard Karls Universit\"{a}t T\"{u}bingen, 72076 T\"{u}bingen, Germany}

\author{Erik Sch\"affer}

\affiliation{Cellular Nanoscience (ZMBP), Eberhard Karls Universit\"{a}t T\"{u}bingen, 72076 T\"{u}bingen, Germany}

\author{Daniel Braun}
\affiliation{Institut f\"{u}r Theoretische Physik, Eberhard Karls Universit\"{a}t T\"{u}bingen, 72076 T\"{u}bingen, Germany}




\date{\today}

\begin{abstract}
We present an innovative optical imaging system for measuring parameters of a small particle {such as a macromolecule or nanoparticle} at the quantum limit of sensitivity. In comparison to the conventional confocal interferometric scattering (iSCAT) approach, {our} setup {adds a second arm to form a Michelson interferometer that allows us to tune a relative phase.} We evaluate the quantum Cramér-Rao bound (QCRB) for different quantum states, including single-mode coherent states, multi-frequency coherent states, and phase-averaged coherent states. Our results show that the proposed setup can achieve the QCRB {of sensitivity} and outperform iSCAT for all considered quantum states for mass and phase estimation of a particle.
\end{abstract}

\maketitle


\textit{Introduction.---}
Recent advancements in optical imaging have significantly improved the estimation of the parameters of the targeted system, and widened applications across fields such as physics, biology, and medicine \cite{taylor2019,taylor2019a,young2018,becker2023,dahmardeh2023,piliarik2014,tsang2016,tsang2019,kose2022,kose2023,lupo2016}. Despite the remarkable progress in classical imaging technologies, even the most advanced conventional microscopes have not yet reached the optimal quantum limit of resolution set by the quantum Cramér-Rao bound (QCRB), which in the context of imaging is the best resolution optimized over all possible measurements and data analysis schemes \cite{helstrom1969,bouchet2021,bouchet2021a,shechtman2014,braunstein1994}. It is given by the inverse of the quantum Fisher information (QFI), which in turn sets an upper bound to the classical Fisher information (CFI) from any specific measurement {to estimate a specific parameter}. Achieving the QCRB would require overcoming all technical problems, such as stability issues thermal and technical noise, leaving only the fundamental quantum noise present in the quantum state of light. Interferometric imaging plays an important role in photometry, where the mass of a small particle, such as a protein molecule, is estimated based on the light it scatters \cite{young2018,priest2019,becker2023,bouchet2021,lin2021,gholamimahmoodabadi2020}. When incident light interacts with a particle it induces a dipole moment leading to Rayleigh scattering \cite{vinogradov2021,bradshaw2020}. The polarizability of the particle is related to its volume, and for a spherical particle gives a linear relationship between mass of the particle and scattered field amplitude \cite{bohren1998}. Due to the small scattering cross-section of individual particles, the estimation of parameters from the particle is challenging. In addition, some light is reflected from the glass surface on which the particle rests.  This reflected field carries no information about the parameters of the particle, thereby reducing the CFI derived from the intensity measurements. {Dark-field microscopy can eliminate this reflected light and the QCRB can be achieved for mass estimation of the nanoparticle if there is no noise from the detector, which, however, is challenging to achieve. {Further, there is no information about the position of the particle or other information carried by the phase  \cite{weigel2014,dong2021}.} In interferometric scattering microscopy (iSCAT), one uses the interference of the scattered light from a nanoparticle and the reflected light from the glass which also allows phase estimation \cite{failla2007,failla2006,taylor2019,dastjerdi2022,young2019}. However, in general iSCAT does not saturate the QCRB for either phase or mass estimation. To address these limitations and understand the fundamental constraints on parameter estimation, we propose a combination of iSCAT with a Michelson interferometer as an alternative and highly flexible setup (MiSCAT, Michelson interferometric scattering microscopy).  We analyze the QFI and the QCRB for this setup and compare it to  
confocal iSCAT \cite{kuppers2023,lindfors2004,hauler2020,hauler2021,wackenhut2020}. It becomes clear that optimizing the phase and amplitude of the light in the second arm of the Michelson interferometer}
is crucial for improving parameter estimation and reaching the quantum limit of sensitivity.

\begin{figure}[t!]
    \centering
    \includegraphics[width=0.8\linewidth]{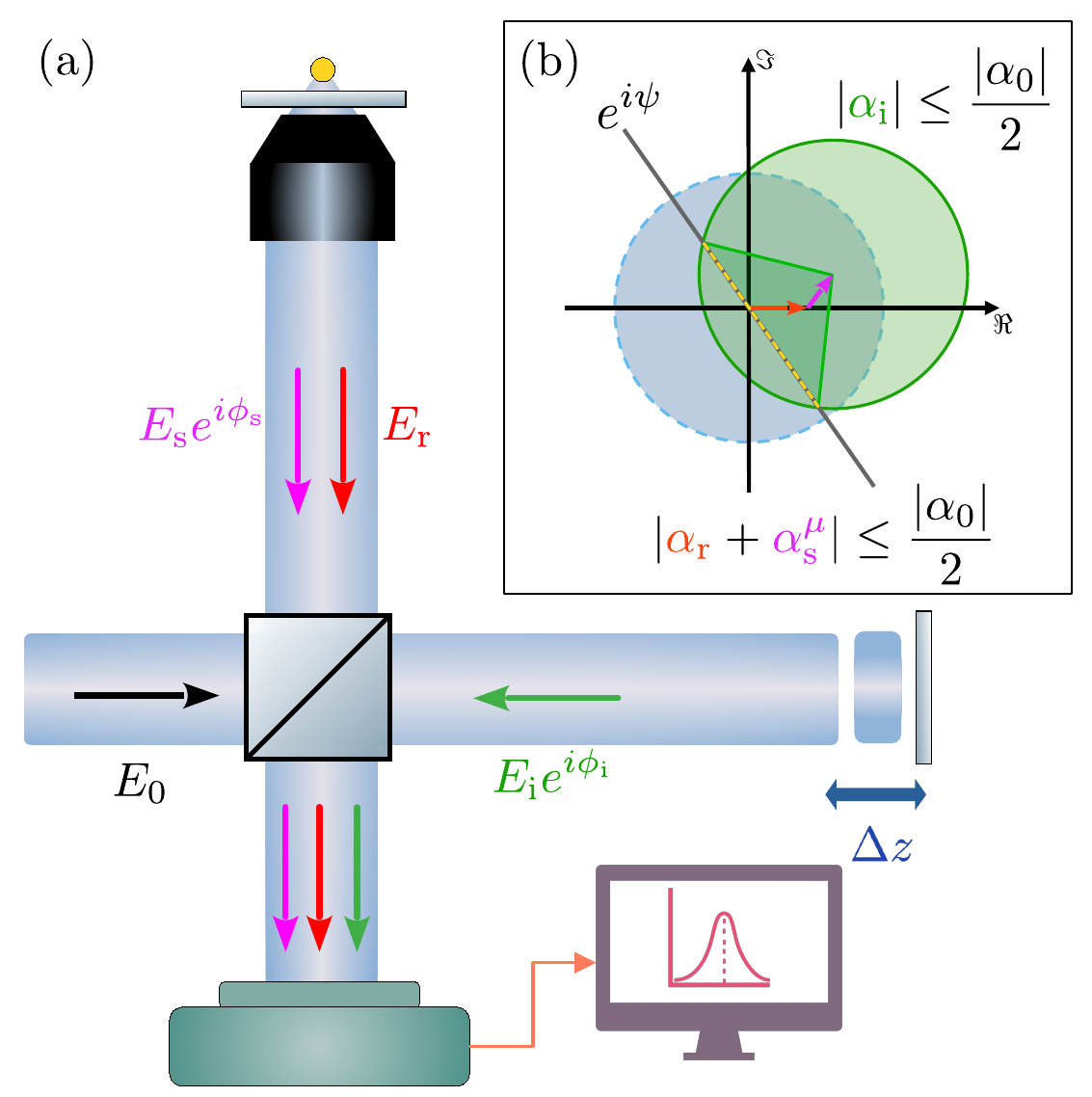}
    \caption{(color online) (a) The setup for MiSCAT mass photometry. A laser source illuminates the sample with an initial incoming field denoted by $E_0$. The light scattered by the particle from the sample is represented by the field $E_{\mathrm{s}}e^{i\phi_\mathrm{s}}$. A first glass plate reflects a portion of the incoming field, denoted by $E_{\mathrm{r}}$. To achieve optimal sensitivity, a phase in the total field adjusted by a second interferometric arm introduces a precisely controlled, $E_{\mathrm{i}}e^{i\phi_\mathrm{i}}$. This additional field is carefully adjusted and ensures that the total field ($E_{\mathrm{r}}+E_{\mathrm{s}}e^{i\phi_\mathrm{s}}+E_{\mathrm{i}}e^{i\phi_\mathrm{i}}$) arriving at the single-point detector gives a specific phase, enabling us to achieve the quantum limit of sensitivity. (b) Corresponding coherent state labels in the complex plane. The yellow dashed line shows the solution space of $\alpha_{\mathrm{i}}$ to saturate the QCRB of mass estimation.}
    \label{fig:Setup-Scheme}
\end{figure}

\textit{Setup.---} {The MiSCAT, shown in Fig.~\ref{fig:Setup-Scheme}(a), is inspired by a nanoscale plasmonic phase sensor recently studied by one of us \cite{hauler2020,wackenhut2020}. Here we optimize it for photometry.} $E_\mathrm{s}$ and $E_\mathrm{r}$ are the real amplitudes of the positive frequency components of the scattered and reflected electric fields for a chosen polarization direction. The first arm of the interferometer contains the scattered light $E_\mathrm{s} e^{i\phi_\mathrm{s}}$ with the parameters imprinted on the quantum state where $\phi_\mathrm{s}$ is the relative phase of the scattered light with respect to the reflected light, and the reflected light $E_\mathrm{r}$ from the glass interface of the sample holder, corresponding to a typical iSCAT configuration \cite{becker2023}. The reference arm (second arm) is used to optimize the total field which includes the imprinted parameter. Thus, we consider the reference arm as adjustable in its length and reflectivity. The adjustable reflectivity allows one to tune the amplitude $E_\mathrm{i}$, while the change of the length of the arm results in a relative phase $\phi_\mathrm{i} = 2 k \Delta z$ compared to the reflected field of the first arm. This yields a positive frequency component of the electric field with fixed polarization direction from the reference arm as $E_\mathrm{i}e^{i \phi_\mathrm{i}}$. Before the detection, a beam splitter superposes the fields from both arms yielding  $E_\mathrm{d} = E_\mathrm{r} + E_\mathrm{s} e^{i \phi_\mathrm{s}} + E_\mathrm{i} e^{i \phi_\mathrm{i}}$ as the positive frequency component of the total field at the detector \cite{hauler2021,wackenhut2020}. We can now investigate the information contained in the scattered light in different kinds of states. We first consider the single-mode states and then multi-frequency mode states.

\textit{Single Mode.---} We consider a single mode coherent state $\rho^0=\ketbra{\alpha_0}{\alpha_0}$ as an initial quantum state produced by a laser source with initial average photon number $\bar n^0 = |\alpha_0|^2$ which translates the positive frequency component of the electric field to labels of the coherent states \cite{glauber1963}. We neglect any non-linear and dissipative effects such that an incoming coherent state propagates to a coherent state of the light at the detector as
\begin{equation}\label{Eq:StateTrasformation}
    | \alpha_0 \rangle \xrightarrow{\mathrm{setup}} \left| \alpha_\mathrm{d} \right\rangle = \left| \alpha_\mathrm{r} + \alpha_\mathrm{s}^\mu + \alpha_\mathrm{i}\right\rangle,
\end{equation}
where the super-script $\mu$ is the parameter to be estimated, carried by the complex label  $\alpha_\mathrm{s}^\mu$. 
The complex labels $\alpha_\mathrm{r}$ and $\alpha_\mathrm{i}$ represent the fields from different arms of the interferometer. In general, the electric field contains both polarization directions in the detection plane. For simplicity, we ignore the second polarization of the field for the coherent state. For a roughly spherical biomolecule with radius much smaller than the wavelength, we have a linear parameter dependence of the label of the coherent state, $\alpha_{\mathrm{s}}^\mu= ms e^{i\phi_{\mathrm{s}}}$ on the mass  $m$, where $\phi_{\mathrm{s}}$ is the phase of the light scattered from the particle, and 
$s$ is a real proportionality factor with dimension of inverse mass. The interesting parameters for this setup are either mass $m$ or phase $\phi_\mathrm{s}$ of the scattered light, $\mu \in \{m, \phi_\mathrm{s}\}$. Since there are no additional photons created in the setup, coherent state labels are constraint by the relation $\left|\alpha_\mathrm{d}\right|^2 \leq |\alpha_0|^2$. This yields an upper bound on each arm, $\left| \alpha_\mathrm{r} + \alpha_\mathrm{s}^\mu \right| \leq{\left| \alpha_0 \right|}/{2}$ and $\left| \alpha_\mathrm{i} \right| \leq {\left| \alpha_0 \right|}/ {2}$. The factor $1/2$ arises from the transformation of the 50:50 beam splitter \cite{fabian2024}. The QFI for the parameter $\mu$ in the coherent state $\rho^\mathrm{d}_\mathrm{c} = |\alpha_\mathrm{d}\rangle \langle\alpha_\mathrm{d}|$ with subscript `c' is given by the partial derivative of $\alpha_\mathrm{s}^\mu$ with respect to the parameter $\mu$ as $\mathscr{F}^{\mu}_\mathrm{c} = 4 \left| \partial_\mu \alpha_{\mathrm{s}}^\mu\right|^2$. It is completely independent of $\alpha_\mathrm{i}$ and $\alpha_\mathrm{r}$. However, we will show later that in general there is no measurement that can achieve the QCRB if we only have $\alpha_\mathrm{r}$ without the reference arm.
Due to the linear parameter dependence on $m$ its QFI becomes 
independent of $m$ and of $\phi_{\mathrm{s}}$ \cite{dong2021}.
The QFI gives a lower bound on the mean square error as $ \delta\mu \geq \sqrt{1/\mathscr{F}^{\mu}_\mathrm{c}} $ \cite{helstrom1969}. {Regarding the number of scattered photons from the particle, $\bar{n}^{\mathrm{s}}_{\mathrm{c}} = |\alpha_\mathrm{s}^\mu|^2$, one obtains a lower bound of the relative uncertainty as}
\begin{eqnarray}
    \frac{\delta m}{m}\sqrt{\bar{n}^{\mathrm{s}}_{\mathrm{c}}} \geq \frac{1}{2}. 
\end{eqnarray}

{For instance, in \cite{becker2023} the number of scattered photons considering a laser with an intensity $I_0=0.1  \ \mathrm{MW}/(\mathrm{cm}^2)$ with wavelength $\lambda = 445  \ \mathrm{nm}$ and effective exposure time $\Delta t=100\ \mathrm{ms}$ from a biomolecule with a mass of $66 \ \mathrm{kDa}$ and a radius 3.8\,nm was found to be $\bar{n}^{\mathrm{s}}_{\mathrm{c}} \approx 220$, which leads to $s\simeq 0.22/$kDa. This leads to a relative uncertainty in the quantum limit for mass estimation of $\delta m \approx 2.22 \ \mathrm{kDa} $, or, keeping exposure time more general, to  $(\delta m)^2 \Delta t \approx 0.5 \ (\mathrm{kDa})^2 /\mathrm{Hz} $.}

The coherent states have definite phase. However, a more realistic model of the electric field produced by a laser contains a randomly fluctuating phase that can be modeled as a mixture of coherent states with uniformly distributed phase, $\rho_{\mathrm{p}}^0 = \int_0^{2\pi} {d\gamma_0} \ketbra{\alpha_0 e^{i\gamma_0}}{\alpha_0 e^{i\gamma_0}}/{2\pi}$, where $\alpha_0$ is a real amplitude and $\gamma_0$ is the corresponding phase that is averaged \cite{allevi2013,mandel1995}. In the detector, we have 
\begin{equation}
    \rho_{\mathrm{p}}^\mathrm{d} =\frac{1}{2\pi} \int_0^{2\pi} d\gamma_0 \left| \alpha_\mathrm{d} e^{i\gamma_0} \right\rangle\!\left\langle\alpha_\mathrm{d} e^{i\gamma_0}  \right|,
\end{equation}
where $\alpha_\mathrm{d} =\alpha_\mathrm{r} + \alpha_\mathrm{s}^\mu + \alpha_\mathrm{i}$, and {subscript `p' stands for the phase averaged coherent state}. Then the QFI for this state becomes
\begin{equation}
    \mathscr{F}^{\mu}_{\mathrm{p}} = 4 \Re^2\left[ \frac{{\alpha}^*_d}{|\alpha_\mathrm{d}|} \partial_\mu \alpha^\mu_{\mathrm{s}} \right] ,
    \label{eq.4}
\end{equation}
where we can define {$\psi = \arg(\partial_\mu\alpha_{\mathrm{s}}^\mu)$ and $\chi = \arg(\alpha_\mathrm{d})$} \cite{fabian2024}. We write the QFI in the form of {$\mathscr{F}^{\mu}_{\mathrm{p}} = 4|\partial_\mu\alpha_{\mathrm{s}}^\mu|^2\cos^2(\psi-\chi)$}. Since $ \cos^2(\psi-\chi)\leq 1$, we see that it is upper bounded by the QFI of the coherent state $\mathscr{F}^{\mu}_{\mathrm{p}}\leq \mathscr{F}^{\mu}_{c}$. Clearly, to estimate a parameter in this phase averaged coherent state, one can first optimize the corresponding phases to reach the QFI of the coherent state. For the parameter '$m$' carried by the field amplitude $\alpha_{\mathrm{s}}$ the QCRB of the relative mass uncertainty becomes
\begin{eqnarray}
     \frac{\delta m}{m}\sqrt{\bar{n}^{\mathrm{s}}_\mathrm{c}} \geq \frac{1}{2|\cos(\psi-\chi)|}.
\end{eqnarray}
This uncertainty implies that it is optimal if the phase $\chi$ of the total field matches the phase $\psi$ of its derivative. {The relation between the field from the first arm and the reference arm of the interferometer is shown in Fig.~\ref{fig:Setup-Scheme}(b) where the coherent state labels are plotted in the complex plain. The blue circle (dashed line) contains all possible coherent state labels $\alpha_\mathrm{r} + \alpha_{\mathrm{s}}^\mu$ of the first arm. The coherent state label $\alpha_\mathrm{i}$ of the reference arm is added to the label of the first arm. Thus all possible coherent states with labels $\alpha_\mathrm{d}$ are located in a circle, centered at the coherent state label of the first arm, which is shown in Fig.~\ref{fig:Setup-Scheme}(b) as green circle (solid line). The superposition yields the total coherent state label at the detector $\alpha_\mathrm{d}(\mu; \alpha_\mathrm{i}, \phi_\mathrm{i})$ as a function of the reference arm's parameters, therefore affecting its phase $\chi(\alpha_\mathrm{i}, \phi_\mathrm{i})$. The coherent state label must have the phase $\psi$ which corresponds to all labels on the {straight} line through the origin. All coherent states with labels in the intersection of this line and the green circle, indicated by the yellow dashed line, saturate the QCRB. In general there is no unique solution, but a continuous set of parameters $\left(|\alpha_\mathrm{i}|,\phi_\mathrm{i}\right)$, saturating the QCRB. The contributions of both arms are bounded by $\left|\alpha_0\right| / 2$ such that the maximum distance of the first arm's coherent state label $\alpha_\mathrm{r} + \alpha_\mathrm{s}^\mu$ to the solution space is $\left|\alpha_0\right|/2$ and, further, that there is always a solution satisfying the condition $\psi = \chi$.}

\textit{Multi frequency-mode.---}Apart from the coherent light, in mass photometry narrow-bandwidth white light is considered in order to avoid speckle effects due to laser light \cite{lindfors2004}. In the following we take the initial state of the field produced by a super-luminescent LED as $\rho_0^{\mathrm{f}}=\ketbra{\{\alpha_0(\omega)\}}{\{\alpha_0(\omega)\}}$, labelled by a set of coherent state amplitudes $\{\alpha_0(\omega)\}$ \cite{glauber1963}. {We assume that for white light {$\alpha_0(\omega) \propto \sqrt{ {1}/{\omega}}$} (flat intensity distribution) in the band width $\Delta\omega$ of interest.} Furthermore, we assume that scattering from the particle does not create any correlations between different frequencies. The resulting state at the detector is then given by $\rho_{\mathrm{f}}^\mathrm{d} =\ketbra{\{\alpha_\mathrm{d}(\omega)\}}{\{\alpha_\mathrm{d}(\omega)\}}$, where $\alpha_\mathrm{d}(\omega)=\alpha_\mathrm{r}(\omega) + \alpha_\mathrm{s}^\mu(\omega) + \alpha_\mathrm{i}(\omega)$ with additional frequency dependency. Since there are no correlations between different modes, we have a continuous product of coherent states, and the  QFI becomes $\mathscr{F}^{\mu}_\mathrm{\mathrm{f}} =4\int_{\Delta \omega} d\omega |\partial_\mu \alpha_{\mathrm{s}}^\mu(\omega)|^2$. For a linear dependency on the mass of the scattered field \cite{becker2023,dong2021}, 
$\alpha_{\mathrm{s}}^\mu(\omega)= ms(\omega) e^{i\phi_{\mathrm{s}}(\omega)}$, we find the lower bound on the relative uncertainty of mass estimation $({\delta m}/{m})\sqrt{\bar{n}^{\mathrm{s}}_{\mathrm{f}}} \geq {1}/{2}$, just as for a single-mode coherent state. The scattered photon number from the particle is given by $\bar{n}^{\mathrm{s}}_{\mathrm{f}}=\int_{\Delta \omega}d \omega |\alpha_{\mathrm{s}}^\mu(\omega)|^2$. If $\bar{n}^{\mathrm{s}}_{\mathrm{f}}= \bar{n}^{\mathrm{s}}_{\mathrm{c}}$, one obtains the same minimal uncertainty for mass estimation as for single-frequency light, $(\delta m)_{\mathrm{c}} = (\delta m)_{\mathrm{f}}$. However, with multi-frequency light with flat distribution over frequencies, typically a much higher total photon number can be achieved than with a single mode, $\bar{n}^{\mathrm{s}}_{\mathrm{f}}\gg \bar{n}^{\mathrm{s}}_{\mathrm{c}}$, which results in much smaller minimal uncertainty for the same sample size, $(\delta m)_{\mathrm{f}} \ll (\delta m)_{\mathrm{c}}$.

Furthermore, we consider a phase averaged multi-frequency initial state, where all frequencies are phase averaged independently with the continuous tensor product over frequencies with a bandwidth $\Delta \omega$. Then the state of the light at the detector becomes 
\begin{equation}
    \rho^\mathrm{d}_{\mathrm{pf}}= \bigotimes_\omega  \int_0^{2\pi}\frac{d\gamma_0(\omega)}{2\pi} \ketbra{\alpha_\mathrm{d}(\omega) e^{i\gamma_0(\omega)}}{\alpha_\mathrm{d}(\omega) e^{i\gamma_0(\omega)}},
\end{equation}
 where `pf' stands for phase averaged multi-frequency coherent state. Assuming no correlations between different frequencies are created by the scattering process, one obtains the QFI as
\begin{equation}
       \mathscr{F}^{\mu}_{\mathrm{pf}} = 4 \int_{\Delta \omega} d\omega \,\Re^2\left[ \frac{{\alpha}^*_d(\omega)}{|\alpha_\mathrm{d}(\omega)|} \partial_\mu \alpha_{\mathrm{s}}^\mu(\omega) \right].
\end{equation}
 Introducing the phases $\psi(\omega) = \arg(\partial_\mu\alpha_{\mathrm{s}}^\mu(\omega))$ and $\chi(\omega) = \arg(\alpha_\mathrm{d}(\omega))$ with frequency dependence, we can rewrite the QFI as $\mathscr{F}^{\mu}_{\mathrm{pf}} = 4 \int_{\Delta \omega} d\omega |\partial_\mu \alpha_{\mathrm{s}}^\mu(\omega)|^2 \cos^2(\psi(\omega)- \chi(\omega))$. Since $\cos^2(\psi(\omega)-\chi(\omega))\leq 1$ for all $\omega$, the QFI of this state is upper bounded by the QFI of the multi-frequency coherent state with the same $\alpha_\mathrm{d}(\omega)$ distribution. With the explicit form of $\alpha_{\mathrm{s}}^\mu(\omega)$ one finds the relative uncertainty in the mass estimation as
\begin{equation}
    \frac{\delta m}{m} \sqrt{\bar{n}^{\mathrm{s}}_{\mathrm{pf}}} \geq \frac{1}{2} \sqrt{\frac{\int_{\Delta \omega} d\omega |s(\omega)|^2}{\int_{\Delta \omega} d\omega |s(\omega)|^2 \cos^2(\psi(\omega)- \chi(\omega))}},
\end{equation}
where $\bar{n}^{\mathrm{s}}_{\mathrm{pf}} = \int_{\Delta \omega} d\omega |\alpha_{\mathrm{s}}^\mu(\omega)|^2$ is the number of scattered photons. This can be optimized and reach the limit of the QFI of the coherent state if for each frequency the phases of the field $\chi(\omega)$ and of its derivative $\psi(\omega)$ match.

\begin{figure*}
    \centering
    \includegraphics[width=\textwidth]{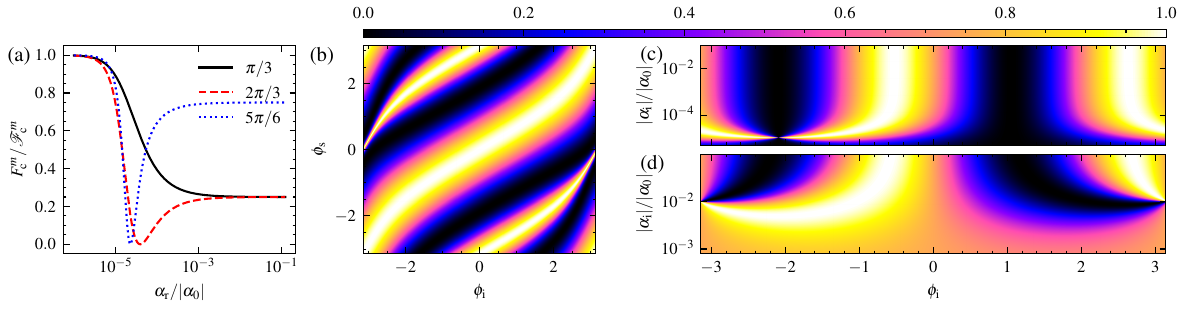}
    \caption{The ratio $F_{\mathrm{c}}^m/\mathscr{F}_{\mathrm{c}}^m $ for coherent states as a function of different parameters considering mass estimation with a scattered field amplitude  $|\alpha_\mathrm{s}| = 2\times10^{-5} |\alpha_0|$. (a) $F_{\mathrm{c}}^m/\mathscr{F}_{\mathrm{c}}^m $ as a function of $\alpha_\mathrm{r}$ in log-scale for the iSCAT considering different phases of the scattered field $\phi_\mathrm{s} \in \{ \pi/3, 2\pi/3,5\pi/6\}$. (b)  Density plot of $F_{\mathrm{c}}^m/\mathscr{F}_{\mathrm{c}}^m $ as a function of $\phi_\mathrm{s}$ and $\phi_\mathrm{i}$ using MiSCAT with $\alpha_\mathrm{r} = 2.3 \times10^{-5} |\alpha_0|$ and $\alpha_\mathrm{i} = 4.5 \times10^{-5} |\alpha_0|$. (c) and (d) the ratio $ F_{\mathrm{c}}^m/\mathscr{F}_{\mathrm{c}}^m $ as a function of $|\alpha_\mathrm{i}|$ and $\phi_\mathrm{i}$ with $\phi_\mathrm{s} = {5\pi}/{6}$ using MiSCAT. In (c) $\alpha_\mathrm{r} = 2.3 \times10^{-5}  |\alpha_0|$ near the value for minimal CFI, and in (d) $\alpha_\mathrm{r} = 10^{-2}  |\alpha_0|$ is used.}
    \label{fig:mass_estimation}
\end{figure*}

\textit{Measurement.---} Typical measurement methods in the experiments of iSCAT include cameras or photo-diodes, both corresponding to an intensity measurement. The positive operator-valued measure (POVM) for the photon number measurement, which for single-mode coherent states also corresponds to intensity measurement, is given by $\Pi_n = \left|n\right\rangle\!\left\langle n\right|$, resulting in a Poisson probability distribution
\begin{equation}
    P(n |\mu) = e^{-|\alpha_\mathrm{d}|^2} \frac{|\alpha_\mathrm{d}|^{2n}}{n!}.
\end{equation}
Then we obtain the CFI, expressed by non-calligraphic $F$, as 
\begin{equation} \label{eq:eq.10}
    F_{\mathrm{c}}^\mu = 4|\partial_\mu\alpha_{\mathrm{s}}^\mu|^2\cos^2(\chi-\psi). 
\end{equation}

{The obtained CFI from a photon number measurement on a single mode coherent state is the same as the QFI of phase-averaged coherent states.} This equivalence arises because the photon number measurement does not capture information about the total phase of the light field. Consequently, the CFI from this measurement is equivalent to the QFI of a phase averaged coherent state. In both cases, achieving the QFI relies heavily on the phase of the reference field. Thus, $\mathscr{F}^{\mu}_{\mathrm{c}}$ is now an upper bound for both the CFI of the coherent state $F_{\mathrm{c}}^\mu$ and the QFI of the phase-averaged state $\mathscr{F}^{\mu}_{\mathrm{p}}$ and $F_{\mathrm{c}}^\mu=\mathscr{F}^{\mu}_{\mathrm{p}}$. 

{In general, the reflected field intensity from the glass interface is three orders of magnitude smaller than the initial field intensity. Additionally, the intensity of scattered light from the particle is approximately six orders of magnitude smaller than the reflected field intensity for a typical iSCAT setup for mass estimation \cite{young2018,becker2023}.} In Fig.~\ref{fig:mass_estimation}(a), we show $F_{\mathrm{c}}^m/\mathscr{F}_{\mathrm{c}}^m $ without reference arm of the interferometer. We fixed the amplitude of the scattered field relative to the initial field as $|\alpha_\mathrm{s}^\mu|=2\times10^{-5} |\alpha_0|$. We see that if there is a reflected field the CFI for estimating the mass from photon number measurements always drops as function of $\alpha_\mathrm{r}$ in a way depending on $\phi_\mathrm{s}$. {The QFI of the coherent state is only saturated for $\alpha_\mathrm{r} = 0$, which corresponds to a dark field microscopy. However, there is always a reflected field in iSCAT setup ($\alpha_\mathrm{r} \neq 0$), preventing the saturation of the QFI.} Furthermore, the reflected field amplitude is usually considerably larger than the scattered field amplitudes and the CFI does not saturate the QCRB. Figs.~\ref{fig:mass_estimation} (b), ~\ref{fig:mass_estimation}(c) and ~\ref{fig:mass_estimation}(d) show the effect of the reference arm using MiSCAT. In Fig.~\ref{fig:mass_estimation}(b), we consider the amplitude of $|\alpha_\mathrm{i}|= 4.5\times10^{-5} |\alpha_0|$ which is large enough to saturate the QCRB for any phase $\phi_\mathrm{s}$ of the scattered field, and show the relation between $\phi_\mathrm{i}$ and $\phi_\mathrm{s}$. In this case of large enough amplitude  $|\alpha_\mathrm{i}|\geq |\alpha_\mathrm{r}+\alpha_\mathrm{s}^\mu|$, we only adjust the phase of the reference arm, and the QFI is reached if $\cos(\psi - \chi) = 0$. In Fig.~\ref{fig:mass_estimation}(c), we consider the parameters of minimum CFI in \ref{fig:mass_estimation}(a), when $\phi_\mathrm{s} = 5\pi/6$ (blue dotted line). In Fig.~\ref{fig:mass_estimation}(d) we consider CFI close to QFI by choosing $|\alpha_\mathrm{r}|=0.01 |\alpha_0|$. In both cases we show the ratio $F_{\mathrm{c}}^m/\mathscr{F}_{\mathrm{c}}^m $ as a function of $|\alpha_\mathrm{i}|$ and $\phi_\mathrm{i}$. If $|\alpha_\mathrm{i}|$ is lower than a certain value we can not reach the QCRB by changing $\phi_\mathrm{i}$. Thus, total field amplitude is not large enough to touch the yellow line in Fig.~\ref{fig:Setup-Scheme}, which is the solution when $\psi = \chi$ for mass estimation. But in the case of $|\alpha_\mathrm{i}|$ large enough to saturate the QFI there are two solutions defined by two different phases $\phi_{\mathrm{i}}$. This implies that a variation of the phase $\phi_{\mathrm{i}}$ for a large enough $\alpha_\mathrm{i}$ allows one to saturate the QCRB with the intensity measurement. The intersection points in Fig.~\ref{fig:mass_estimation}(c) and ~\ref{fig:mass_estimation}(d) correspond to a vanishing field, $\alpha_\mathrm{d}(\mu; \alpha_\mathrm{i}, \phi_\mathrm{i}) = 0$, i.e.~a vacuum state. Thus, the CFI is not defined at this point since the vacuum has no defined phase. {When the phase shift due to the particle is $\phi_\mathrm{s} = \pi/2$, the CFI and signal-to-noise ratio (SNR) for iSCAT becomes extremely small for mass estimation \cite{fabian2024}. This implies that the mean value of the probability distribution after the intensity measurement does not change with respect to the parameter `$m$'. In \cite{fabian2024}, we consider this case and compare it to MiSCAT with optimal phase $\phi_\mathrm{i}$ and amplitude $\alpha_\mathrm{i}$. Then, the mean value of the optimal probability distribution including the second arm becomes sensitive to the parameter `$m$'. This is corroborated by the SNR for mass estimation which can be maximized by correctly tuning the phase $\phi_\mathrm{i}$ of the second arm of the interferometer.}

\begin{figure}[t!]
    \centering
    \includegraphics[width=\linewidth]{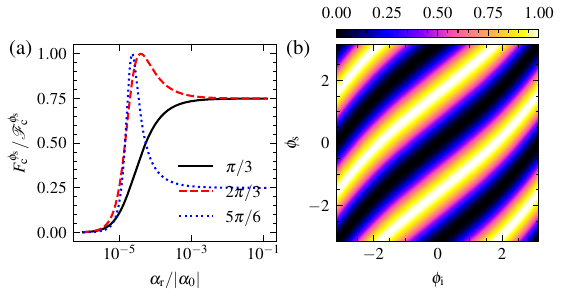}
    \caption{The ratio $F_{\mathrm{c}}^{\phi_\mathrm{s}}/\mathscr{F}_{\mathrm{c}}^{\phi_\mathrm{s}} $ for coherent states as a function of different parameters considering phase estimation with the scattered field $|\alpha_\mathrm{s}| = 2\times 10^{-5} |\alpha_0|$. (a) The ratio $F_{\mathrm{c}}^{\phi_\mathrm{s}}/\mathscr{F}_{\mathrm{c}}^{\phi_\mathrm{s}} $  as a function of $|\alpha_\mathrm{r}|$ in log-scale for the iSCAT considering different phases of the scattered field $\phi_\mathrm{s} \in \{ \pi/3, 2\pi/3,5\pi/6\}$. (b) Density plot of the ratio $F_{\mathrm{c}}^{\phi_\mathrm{s}}/\mathscr{F}_{\mathrm{c}}^{\phi_\mathrm{s}} $ as a function of $\phi_\mathrm{s}$ and $\phi_\mathrm{i}$ using MiSCAT with $\alpha_\mathrm{r} = 0.01|\alpha_0|$ and $\alpha_\mathrm{i} = 0.045 |\alpha_0|$.}
    \label{fig:phase-estimation}
\end{figure}

{One may also be interested in estimating the phase of the scattered field. The estimation of the phase can contain additional information, e.g.~in a wide-field iSCAT setup, the phase is linked to the position of the particle \cite{dong2021}. There is a $\pi/2$ difference between optimally estimating $\phi_\mathrm{s}$ and $m$,
  which arises from an additional factor '$i$' in the derivative of $ \alpha_{\mathrm{s}}^\mu$ with respect to $\phi_\mathrm{s}$ in Eq.~(\ref{eq.4}).} In Fig.~\ref{fig:phase-estimation}(a), we show the ratio ${F^{\phi_\mathrm{s}}_\mathrm{c}}/{\mathscr{F}^{\phi_\mathrm{s}}_\mathrm{c}}$ as a function of $\alpha_\mathrm{r}$ without the reference arm of the interferometer. Increasing $\alpha_\mathrm{r}$ increases ${F^{\phi_\mathrm{s}}_\mathrm{c}}/{\mathscr{F}^{\phi_\mathrm{s}}_\mathrm{c}}$, which is a behavior opposite to the one for mass estimation. If $\alpha_\mathrm{r}=0$, the intensity measurement does not contain any phase information about $\alpha_\mathrm{s}$ \cite{hauler2020}. Thus, the existence of a relative phase and hence its measurability hinges on a non-vanishing value of $\alpha_\mathrm{r}$. However, to saturate the QCRB requires an $|\alpha_\mathrm{i}|$ larger than $|\alpha_\mathrm{r}+ \alpha_\mathrm{s}^{\mu}|$. We see in Fig.~\ref{fig:phase-estimation}(b) for a small $\alpha_\mathrm{r}=0.01 |\alpha_0|$ that we can saturate the QFI $\mathscr{F}_\mathrm{c}^{\phi_\mathrm{s}}$ of a coherent state for the phase estimation as well, if $\alpha_\mathrm{i}$ is large enough and the phase $\phi_\mathrm{i}$ correctly tuned. {In \cite{fabian2024}, we also show the relation between the signal-to-noise ratio (SNR) considering both cases, for iSCAT and MiSCAT, as $\phi_\mathrm{s}\rightarrow 0$. We see that the SNR scales quadratic $\sim \phi_\mathrm{s}^2$ for iSCAT and linearly $\sim\phi_\mathrm{s}$ for MiSCAT for estimating $\phi_\mathrm{s}$. Thus, one can optimize the SNR by adjusting the phase $\phi_\mathrm{i}$ of the second arm and achieve a better estimation of the phase shift due to the particle when it is very small.}

In the case of a multi-frequency coherent state, we can replace the Poisson probability distribution with a probability density for each frequency. Since we do not have any correlation between different frequency modes, the CFI for multi-frequency coherent state in Eq. (\ref{eq:eq.10}) generalizes to 
\begin{eqnarray}
    {F}^\mu_{\mathrm{f}} = 4 \int_{\Delta \omega} d\omega |\partial_\mu \alpha_{\mathrm{s}}^\mu(\omega)|^2 \cos^2(\psi(\omega)- \chi(\omega)).
\end{eqnarray}
Considering ${F}^{\mu}_{\mathrm{f}}=\mathscr{F}^{\mu}_{\mathrm{pf}}$, the QFI $\mathscr{F}^{\mu}_{\mathrm{f}}$  for the multi-frequency coherent state is an upper bound for both the CFI $F_{\mathrm{f}}^\mu$ for multi-frequency photon number measurement on coherent states and the QFI $\mathscr{F}^{\mu}_{\mathrm{pf}}$ for multi-frequency phase averaged states. The saturation of the $\mathscr{F}^{\mu}_{\mathrm{f}}$ highly depends on the correct phase of the total field. In the experiment, if one can not eliminate the reflected field, it is necessary to adjust the total phases of the field $\chi(\omega)$ of the detection modes using the reference arm of the interferometer to match with the phase $\psi(\omega)$ of the derivative for each frequency component with respect to the parameter to be estimated. 

\textit{Conclusion.---} We proposed a novel setup for optical photometry based on Rayleigh light scattering by combining confocal microscopy with a Michelson interferometer and showed how to optimize the setup for maximum sensitivity by varying the phase and amplitude of the electric field of the reference arm. We analyzed the QFI and QCRB for various initial states of the incoming field, including single-mode coherent states, multifrequency coherent states, and phase-averaged coherent states. The classical CRB for photon number measurement saturates the QCRB using MiSCAT, which can outperform iSCAT for estimation of both mass and phase shift due to the particle if the phase and amplitude of the reference field are adjusted appropriately.

\textit{Acknowledgment.---} We thank Anita Jannasch and Viktor Schiff for the invaluable insights and discussions, which significantly enriched this work.

\nocite{*}
\bibliographystyle{apsrev4-2}
\bibliography{apssamp_emre}

\end{document}


\preprint{APS/123-QED}

\title{Pushing the Boundaries: Interferometric Mass Photometry at the Quantum Limit of Sensitivity}

\author{Fabian M\"uller}
\affiliation{Institut f\"{u}r Theoretische Physik, Eberhard Karls Universit\"{a}t T\"{u}bingen, 72076 T\"{u}bingen, Germany}

\author{Emre K\"ose}
\affiliation{Institut f\"{u}r Theoretische Physik, Eberhard Karls Universit\"{a}t T\"{u}bingen, 72076 T\"{u}bingen, Germany}

\author{Alfred J. Meixner}
\affiliation{Institut f\"{u}r Physikalische und Theoretische Chemie, Eberhard Karls Universit\"{a}t T\"{u}bingen, 72076 T\"{u}bingen, Germany}

\author{Erik Sch\"affer}

\affiliation{Cellular Nanoscience (ZMBP), Eberhard Karls Universit\"{a}t T\"{u}bingen, 72076 T\"{u}bingen, Germany}

\author{Daniel Braun}
\affiliation{Institut f\"{u}r Theoretische Physik, Eberhard Karls Universit\"{a}t T\"{u}bingen, 72076 T\"{u}bingen, Germany}

\date{\today}

\begin{abstract}

\end{abstract}

\maketitle


\onecolumngrid

\section{Supplemental Material}

\subsection{Field Scattering from MiSCAT Setup}
The initial field is $\alpha_0$ on the first mode and vacuum on the second mode. The action of the setup on this state is 
\begin{align}
    \left(\alpha_0, 0\right) &\xrightarrow{\mathrm{bs}} \left( \frac{\alpha_0}{\sqrt{2}}, \frac{\alpha_0}{\sqrt{2}} \right) \xrightarrow{\mathrm{arms}} \left( \tilde\alpha_\mathrm{r} + \tilde\alpha_{\mathrm{s}}^\mu, \tilde\alpha_{\mathrm{i}} \right),\\
    &\xrightarrow{\mathrm{bs}} \left( \dots, \frac{1}{\sqrt{2}} \left( \tilde\alpha_\mathrm{r} + \tilde\alpha_{\mathrm{s}}^\mu + \tilde\alpha_{\mathrm{i}} \right)\right) = \left( \dots, \alpha_\mathrm{r} + \alpha_{\mathrm{s}}^\mu + \alpha_{\mathrm{i}} \right),
\end{align}
with the redefined amplitudes \(\alpha_\mathrm{r,s,i} := \frac{\tilde\alpha_\mathrm{r,s,i}}{\sqrt{2}}\). As there is no additional photon source in the setup, before the recombination of the two arms by the beam splitter (bs), it holds \(\left|\tilde\alpha_\mathrm{r} + \tilde\alpha_{\mathrm{s}}^\mu \right| \leq \frac{\left|\alpha_0\right|}{\sqrt{2}}\) and \(\left|\tilde\alpha_{\mathrm{i}}\right| \leq \frac{\left|\alpha_0\right|}{\sqrt{2}}\). Thus for the redefined amplitudes holds \(\left|\alpha_\mathrm{r} + \alpha_{\mathrm{s}}^\mu \right| = \left|\frac{\tilde\alpha_\mathrm{r} + \tilde\alpha_{\mathrm{s}}}{\sqrt{2}}\right| \leq \frac{\left|\alpha_0\right|}{2}\) and likewise \(\left|\alpha_{\mathrm{i}}\right| \leq \frac{\left|\alpha_0\right|}{2}\)

\subsection{Quantum Cramer Rao Bound and Quantum Fisher Information}
\subsubsection{Coherent States} For a given quantum state, the QFI is given by 
\begin{equation} \label{eq:Defthe QFI}
\mathscr{F}^{\mu}=\operatorname{Tr}\left(\rho \mathscr{L}^{2}\right),
\end{equation}
where $\mathscr{L}$ is the symmetric logarithmic derivative of $\rho_{\mu}$ defined as 
\begin{equation}
\rho_{\mathrm{\mu}}\mathscr{L}+\mathscr{L} \rho_{\mathrm{\mu}}=2 \partial_{\mu} \rho_{\mathrm{\mu}}.
\end{equation}
In case of coherent states the QFI becomes
\begin{equation}
\mathscr{F}^\mu_c=4 \sum_{k}\left(\left\langle\partial_{\mu} \alpha_{k} \mid \partial_{\mu} \alpha_{k}\right\rangle-\left|\left\langle\partial_{\mu} \alpha_{k} \mid \alpha_{k}\right\rangle\right|^{2}\right).
\end{equation}
In the Fock basis one can write
\begin{equation} \label{eq:coherentstatefock}
\left|\alpha_{k}\right\rangle=e^{-\left|\alpha_{k}\right|^{2} / 2} \sum_{n=0}^{\infty} \frac{\alpha_{k}^{n}}{(n !)^{1 / 2}}|n\rangle_k.
\end{equation}
Calculating the first and second term in the QFI using the orthonormality of Fock states we have
\begin{equation}
\begin{split}
\left\langle\partial_{\mu} \alpha_{k} \mid \partial_{\mu} \alpha_{k}\right\rangle &=\left|\partial_{\mu} \alpha_{k}\right|^{2}\left[\left|\alpha_{k}\right|^{2} e^{-\left|\alpha_{k}\right|^{2}} \sum_{n=0}^{\infty} \frac{\left|\alpha_{k}\right|^{2 n}}{n !}-\left|\alpha_{k}\right| \alpha_{k} e^{-\left|\alpha_{k}\right|^{2}} \sum_{n=1}^{\infty} \frac{\left|\alpha_{k}\right|^{2(n-1)}}{(n-1) !}\right.\\&\left.-\left|\alpha_{k}\right| \alpha_{k}^{*} e^{-\left|\alpha_{k}\right|^{2}} \sum_{n=1}^{\infty} \frac{\left|\alpha_{k}\right|^{2(n-1)}}{(n-1) !}+e^{-\left|\alpha_{k}\right|^{2}} \sum_{n=1}^{\infty} \frac{n\left|\alpha_{k}\right|^{2(n-1)}}{(n-1) !}\right],
\end{split}
\end{equation}
and 
\begin{equation}
\left\langle\partial_{\mu} \alpha_{k} \mid \alpha_{k}\right\rangle=\partial_{\mu} \alpha_{k}^{*}\left[-\left|\alpha_{k}\right| e^{-\left|\alpha_{k}\right|^{2}} \sum_{n=0}^{\infty} \frac{\left|\alpha_{k}\right|^{2 n}}{n !}+\alpha_{k} e^{-\left|\alpha_{k}\right|^{2}} \sum_{n=1}^{\infty} \frac{\left|\alpha_{k}\right|^{2(n-1)}}{(n-1) !}\right].
\end{equation}
Using the following properties of the exponential series, $\sum_{n=0}^{\infty} \frac{\left|\alpha_{k}\right|^{2 n}}{n !}=e^{\left|\alpha_{k}\right|^{2}}$ and $\sum_{n=0}^{\infty} \frac{(n+1)\left|\alpha_{k}\right|^{2 n}}{n !}=\left(1+\left|\alpha_{k}\right|^{2}\right) e^{\left|\alpha_{k}\right|^{2}}$, we can simplify 
\begin{equation}
\begin{split}
    \left\langle\partial_{\mu} \alpha_{k} \mid \partial_{\mu} \alpha_{k}\right\rangle&=\left|\partial_{\mu} \alpha_{k}\right|^{2}\left(2\left|\alpha_{k}\right|^{2}-\left|\alpha_{k}\right| \alpha_{k}-\left|\alpha_{k}\right| \alpha_{k}^{*}+1\right) \\\left|\left\langle\partial_{\mu} \alpha_{k} \mid \alpha_{k}\right\rangle\right|^{2} &=\left|\partial_{\mu} \alpha_{k}\right|^{2}\left(2\left|\alpha_{k}\right|^{2}-\left|\alpha_{k}\right| \alpha_{k}-\left|\alpha_{k}\right| \alpha_{k}^{*}\right).
\end{split}
\end{equation}
Finally the QFI \cite{bouchet2021} for the multimode coherent state becomes
\begin{equation}
\mathscr{F}^{\mu}_c=4 \sum_{k=1}\left|\partial_{\mu} \alpha_{k}\right|^{2}.
\label{eq2}
\end{equation}
We give the general formula for multimode coherent state considering single parameter estimation. In case of a single mode with amplitude $ \alpha_\mathrm{d} = \alpha_\mathrm{r} + \alpha_\mathrm{s}^\mu + \alpha_\mathrm{i}$ on the detector, one can find the QFI $\mathscr{F}_c^{\mu}=4 \left|\partial_{\mu} \alpha_\mathrm{s}^{\mu}\right|^{2}$. 

\subsubsection{Multi-Frequency Coherent States}
To derive the form of the QFI for a continuous frequency interval, we start with a quantized frequency space. The QFI for multimode coherent state in discrete form is given by Eq.~\eqref{eq2}. This state has no correlations between different frequencies. In the continuum limit one obtains
\begin{equation}
    \mathscr{F}_\mathrm{f}^\mu \xrightarrow{L \rightarrow \infty} 4 \int d\omega \, |\partial_\mu \tilde\alpha_{\mathrm{s}}^\mu(\omega)|^2,
\end{equation}
with a photon number density $n^\mathrm{f}(\omega) = |\tilde{\alpha}(\omega)|^2$, where $\tilde{\alpha}(\omega)$ is in the dimension of $1/\sqrt{\omega}$. In the main text we omit the tilde from the coherent state labels.

\subsubsection{Phase Averaged Coherent States}
The phase averaged coherent state is defined as 
\begin{equation}
    \rho = \int_0^{2 \pi} \frac{d\gamma_0}{2 \pi} \ketbra{\alpha e^{i\gamma_0}}{\alpha e^{i\gamma_0}}.
\end{equation}
By inserting the explicit form of the coherent states Eq.~\eqref{eq:coherentstatefock} and usage of $\int_0^{2 \pi} d\phi e^{i \phi (n - m)} = 2 \pi \delta_{nm}$, one obtains
\begin{equation}
    \rho = \sum_{n,m} \int_0^{2 \pi} \frac{d\gamma_0}{2 \pi} e^{-|\alpha|^2} \frac{\alpha^n {\alpha^\ast}^m e^{i \gamma_0 (n - m)}}{(n!)^{1/2} (m!)^{1/2}} \ketbra{n}{m} = \sum_{n} e^{-|\alpha|^2} \frac{|\alpha|^{2 n}}{n!} \ketbra{n}{n} = \sum_{n} P_n \ketbra{n}{n}.
\end{equation}
This shows that phase averaged states are diagonal in the Fock basis, distributed with Poisson's distribution $P_n$. The SLD of this state is directly obtained from its definition $\partial \rho = \frac{1}{2}(\mathscr{L}\rho + \rho\mathscr{L})$ as
\begin{equation}
    \mathscr{L} = \sum_n \frac{\partial_\mu P_n}{P_n} \ketbra{n}{n} = -2 \Re{[\alpha^\ast \partial_\mu \alpha]} \sum_n \left( 1 - \frac{n}{|\alpha|^2}\right) \ketbra{n}{n}.
\end{equation}
In the second step we used that the derivative of the Poisson distribution is 
\begin{equation}
    \partial_\mu P_n = -2 \Re{[\alpha^\ast \partial_\mu \alpha]} \left( 1 - \frac{n}{|\alpha|^2}\right) P_n.
\end{equation}
Then the QFI becomes
\begin{equation}
    \mathscr{F^\mu_p} = 4 \Re^2{[\alpha^\ast \partial_\mu \alpha]} \sum_n \left( 1 - \frac{n}{|\alpha|^2}\right)^2 P_n = 4 \Re^2{[\alpha^\ast \partial_\mu \alpha]} \left(1 - \frac{2}{|\alpha|^2} \langle n \rangle + \frac{1}{|\alpha|^4} \langle n^2 \rangle \right),
\end{equation}
from its definition in Eq.~\eqref{eq:Defthe QFI}. The expectation values occurring in the above equation of the QFI are $\langle n \rangle = |\alpha|^2$ and $\langle n^2 \rangle = |\alpha|^4 + |\alpha|^2$ for a Poisson's distribution. Inserting them into the QFI yields
\begin{equation}
    \mathscr{F^\mu_p} = 4 \frac{\Re^2{[\alpha^\ast \partial_\mu \alpha]}}{|\alpha|^2} = 4 \Re^2{\left[\frac{\alpha^\ast}{|\alpha|} \partial_\mu \alpha\right]}.
\end{equation}
The polar representation of the coherent state labels $\alpha = |\alpha| e^{i \chi}$ and its derivative $\partial_\mu \alpha = |\partial_\mu \alpha| e^{i \psi}$ simplifies the QFI to
\begin{equation}
    \mathscr{F^\mu_p} = 4 |\partial_\mu \alpha|^2 \Re^2{\left[ e^{-i \chi} e^{i \psi} \right]} = 4 |\partial_\mu \alpha|^2 \Re^2{\left[ e^{i (\psi - \chi)} \right]} = 4 |\partial_\mu \alpha|^2 \cos^2{\left( \psi - \chi \right)}.
\end{equation}
Thus, one finds $\mathscr{F}^\mu_p = \mathscr{F}^\mu_c \cos^2{\left( \psi - \chi \right)} \leq \mathscr{F}^\mu_c$. As $\cos^2{x} \leq 1$ the QFI for phase averaged states is upper bounded by the QFI of a coherent state.

\subsubsection{Multi-Frequency Phase-Averaged Coherent States}
The result from the previous section is converted to the multi-frequency domain as before. The state is defined as 
\begin{equation}
    \rho = \bigotimes_\omega \int_0^{2\pi} \frac{d\gamma_0(\omega)}{2\pi} \ketbra{\alpha(\omega) e^{i \gamma_0(\omega)}}{\alpha(\omega) e^{i \gamma_0(\omega)}},
\end{equation}
where $\otimes_\omega$ stands for contunious tensor product, for white light $\alpha(\omega) \propto \sqrt{\frac{1}{\omega}}$ (flat intensity distribution). As each single frequency is phase averaged, there are no correlation between the frequencies. Thus the QFI for this state becomes
\begin{equation}
    \mathscr{F} = 4 \int_{\Delta \omega} d\omega \, \Re^2\left[ \frac{\alpha^\ast(\omega)}{|\alpha(\omega)|} \partial_\mu \alpha(\omega) \right].
\end{equation}
corresponding to the QFI of a single frequency coherent state. 

\subsection{Measurement to Estimate Mass}
In iSCAT experiments, the standard detection methods are either cameras or photodiodes, both of which record intensity. For photon number measurements, the positive operator-valued measure (POVM) is represented as $\Pi_n = \left|n\right\rangle\!\left\langle n\right|$, which corresponds to the intensity measurement in the case of single-mode coherent states and results in a Poisson probability distribution given by 
\begin{equation}
    P(n |\mu) = e^{-|\alpha_\mathrm{d}|^2} \frac{|\alpha_\mathrm{d}|^{2n}}{n!},
\end{equation}
where $|\alpha_\mathrm{d}|^2$ is the mean number of photons at the detector. As given in the main text, we have $\alpha_\mathrm{d}=\alpha_\mathrm{r} + \alpha_\mathrm{s}^m + \alpha_\mathrm{i}$ for MiSCAT, and we consider $\alpha_{\mathrm{s}}^m= ms e^{i\phi_{\mathrm{s}}}$ as the field scattered from the particle, which carries the information about the mass $m$ and the phase $\phi_\mathrm{s}$ of the particle. In this case, we are interested in estimating mass and we set $\mu=m$. Usually $|\alpha_\mathrm{d}|^2$ is very large for a typical setup. One can approximate the Poisson probability distribution by a Gaussian probability distribution using the well known Stirling formula to get
\begin{equation}
    P(n |\mu=m) \approx\frac{\exp\left(\frac{-(n-(|\alpha_\mathrm{d}|^2-\frac{1}{2}))^2}{2|\alpha_\mathrm{d}|^2} \right) }{|\alpha_\mathrm{d}|\sqrt{2\pi}}.
\end{equation}

\begin{figure}
    \centering
    \includegraphics[width=0.7\linewidth]{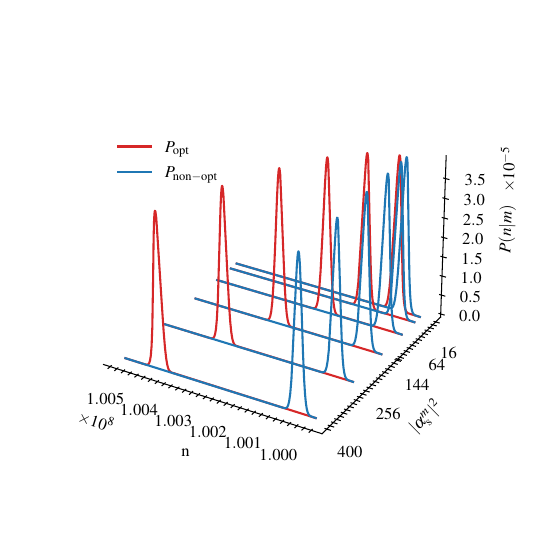}
    \caption{(color online) The photon number distribution $ P(n |\mu=m)$ as a function of $|\alpha_{\mathrm{s}}^m|^2$ for optimized (red) and non-optimized (blue) probability distributions to estimate mass}
    \label{fig:massPn}
\end{figure}

In Fig.\ref{fig:massPn} we show the probability distribution as a function of $n$ and $|\alpha_\mathrm{s}^m|^2$.  When $\phi_\mathrm{s}=\pi/2$, the signal-to-noise ratio (SNR) for iSCAT is extremely small. Then we include the second arm of the interferometer by setting $|\alpha_\mathrm{r} |=|\alpha_\mathrm{i}|$ and optimizing $P(n |\mu=m)$ over $\phi_\mathrm{i}$ to estimate the mass using MiSCAT. One can see that the mean value of the probability distribution using MiSCAT, in red, optimized to saturate the QFI for mass estimation changes when one change $|\alpha_\mathrm{s}^m|^2$. However, it remains constant for a non-optimized probability distribution using iSCAT, in blue, which means that for $\phi_\mathrm{s}=\pi/2$ one cannot estimate the mass with iSCAT. 

\subsection{Signal To Noise Ratio}
While the most fundamental analysis of the sensitivity of any measurement device is based on the QFI and QCRB, similar conclusions how MiSCAT improves the sensitivity compared to iSCAT can be drawn from a simple analysis of the signal-to-noise ratio. We have a Poisson distribution of the photon number after the intensity measurement with a mean value given by the intensity of the total field  $I_\mathrm{d}$ at the detector. We define the signal as the change in the mean $ I_{\mathrm{d}} $ with respect to the parameter of interest $\mu$ as $\mu \frac{\partial I_{\mathrm{d}}}{\partial \mu}$. The noise is represented by the standard deviation due to shot noise, given by $\sqrt{I_{\mathrm{d}}}$. We begin with the iSCAT setup, considering the electric field at the detector without the second arm. The electric field at the detector can be expressed as $
E_{\mathrm{d}} = E_{\mathrm{r}} + E_{\mathrm{s}} e^{i \phi_{\mathrm{s}}},$
leading to the intensity 
\begin{equation}
   I_\mathrm{d}^{(1)}= E_{\mathrm{r}}^{2} + 2 E_{\mathrm{r}} E_{\mathrm{s}} \cos{\left(\phi_{\mathrm{s}} \right)} + E_{\mathrm{s}}^{2}.
\end{equation}
The mass `$m$' linearly scales with $E_\mathrm{s}$ and it gives the signal $2 E_{\mathrm{r}} E_{\mathrm{s}} \cos{\left(\phi_{\mathrm{s}} \right)} $  and $E_{\mathrm{s}}\ll E_{\mathrm{r}}$. Thus, the SNR \cite{dong2021} becomes
\begin{equation}
    \mathrm{SNR}_1(m) = \frac{ 2 E_{\mathrm{r}} E_{\mathrm{s}} \cos{\left(\phi_{\mathrm{s}} \right)}}{\sqrt{E_{\mathrm{r}}^{2} + 2 E_{\mathrm{r}} E_{\mathrm{s}} \cos{\left(\phi_{\mathrm{s}} \right)} + E_{\mathrm{s}}^{2}}}.
\end{equation}

We are also interested in the parameter \( \phi_{\mathrm{s}} \). To understand the SNR around $ \phi_{\mathrm{s}} \sim 0 $, we can expand the cosine terms up to the second order, $
\cos(\phi_{\mathrm{s}}) \approx 1 - \frac{\phi_{\mathrm{s}}^2}{2}
$, and find
\begin{equation}
    I_\mathrm{d}^{(1)}\approx E_{\mathrm{r}}^{2} - E_{\mathrm{r}} E_{\mathrm{s}} \phi_{\mathrm{s}}^{2} + 2 E_{\mathrm{r}} E_{\mathrm{s}} + E_{\mathrm{s}}^{2}.
\end{equation}
We observe that the first term involving $ \phi_{\mathrm{s}} $ is quadratic. Therefore, we can consider the signal related to the parameter as \( 2 \phi_{\mathrm{s}}^2 E_{\mathrm{r}} E_{\mathrm{s}} \), which scales $ \propto\phi_{\mathrm{s}}^2 $. Consequently, the SNR for iSCAT can be expressed for small values of $ \phi_{\mathrm{s}} $ as
\begin{equation}
    \mathrm{SNR}_1(\phi_\mathrm{s})=\frac{2 \phi_\mathrm{s}^2 E_{\mathrm{r}} E_{\mathrm{s}} }{\sqrt{E_{\mathrm{r}}^{2} + 2 E_{\mathrm{r}} E_{\mathrm{s}} + E_{\mathrm{s}}^{2}}}.
\end{equation}

Let us now include the second arm of the interferometer as proposed in the main text for MiSCAT. Considering the total electric field at the detector as 
$ E_{\mathrm{d}} = E_{\mathrm{r}} + E_{\mathrm{s}} e^{i \phi_{\mathrm{s}}} + E_{\mathrm{i}} e^{i \phi_{\mathrm{i}}},
$ one can find the total intensity as
\begin{equation}
    I_\mathrm{d}^{(2)} =E_{\mathrm{i}}^{2} + 2 E_{\mathrm{i}} E_{\mathrm{r}} \cos{\left(\phi_{\mathrm{i}} \right)} + 2 E_{\mathrm{i}} E_{\mathrm{s}} \cos{\left(\phi_{\mathrm{i}} - \phi_{\mathrm{s}} \right)} + E_{\mathrm{r}}^{2} + 2 E_{\mathrm{r}} E_{\mathrm{s}} \cos{\left(\phi_{\mathrm{s}} \right)} + E_{\mathrm{s}}^{2}.
\end{equation}
We have two terms that are linear in $ E_{\mathrm{s}} $, which translate into the signal used to analyze mass. Therefore, we can express the SNR for MiSCAT, including the second arm, as
\begin{equation}
    \mathrm{SNR}_2(m) = \frac{2 E_{\mathrm{r}} E_{\mathrm{s}} \cos{\left(\phi_{\mathrm{s}} \right)}+ 2 E_{\mathrm{i}} E_{\mathrm{s}} \cos{\left(\phi_{\mathrm{i}}- \phi_{\mathrm{s}} \right)}}{\sqrt{E_{\mathrm{i}}^{2} + 2 E_{\mathrm{i}} E_{\mathrm{r}} \cos{\left(\phi_{\mathrm{i}} \right)} + 2 E_{\mathrm{i}} E_{\mathrm{s}} \cos{\left(\phi_{\mathrm{i}} - \phi_{\mathrm{s}} \right)} + E_{\mathrm{r}}^{2} + 2 E_{\mathrm{r}} E_{\mathrm{s}} \cos{\left(\phi_{\mathrm{s}} \right)} + E_{\mathrm{s}}^{2}}}.
\end{equation}

In the case of estimating a small phase shift $ \phi_\mathrm{s} \sim 0 $, we have by expanding the cosine term up to first order
\begin{equation}
    I_\mathrm{d}^{(2)}\approx E_{\mathrm{i}}^{2} + 2 E_{\mathrm{i}} E_{\mathrm{r}} \cos{\left(\phi_{\mathrm{i}} \right)} + 2 E_{\mathrm{i}} E_{\mathrm{s}} \phi_{\mathrm{s}} \sin{\left(\phi_{\mathrm{i}} \right)} + 2 E_{\mathrm{i}} E_{\mathrm{s}} \cos{\left(\phi_{\mathrm{i}} \right)} + E_{\mathrm{r}}^{2} + 2 E_{\mathrm{r}} E_{\mathrm{s}} + E_{\mathrm{s}}^{2}.
\end{equation}
The signal is given by the term $2 E_{\mathrm{i}} E_{\mathrm{s}} \phi_{\mathrm{s}} \sin{\left(\phi_{\mathrm{i}} \right)}$, while the remaining terms contribute to the noise. By assuming $E_{\mathrm{s}}\ll E_{\mathrm{r}}$ and keeping only the first-order term in $E_{\mathrm{s}}$, we can express the signal-to-noise ratio as
\begin{equation}
    \mathrm{SNR}_2(\phi_\mathrm{s})=\frac{2 E_{\mathrm{i}} E_{\mathrm{s}} \phi_\mathrm{s}\sin{\left(\phi_{\mathrm{i}} \right)}}{\sqrt{E_{\mathrm{i}}^{2} + 2 E_{\mathrm{i}} E_{\mathrm{r}} \cos{\left(\phi_{\mathrm{i}} \right)} + E_{\mathrm{r}}^{2}}}.
\end{equation}

\begin{figure}%
    \centering
    \subfloat[\centering The $\mathrm{SNR}(m)$ as a function of $\phi_\mathrm{i}$.\label{fig:snr1}]{{\includegraphics[width=8.5cm]{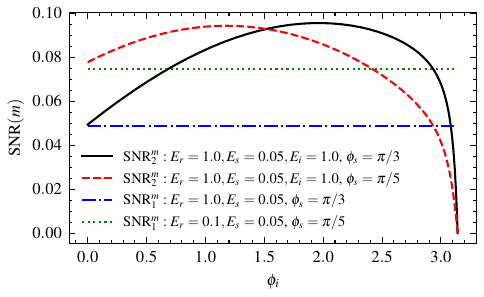} }}%
    \qquad
    \subfloat[\centering The $\mathrm{SNR}(\phi_\mathrm{s})$ in logarithmic scale as a function of $\phi_{\mathrm{i}}$.\label{fig:snr2}]{{\includegraphics[width=8.5cm]{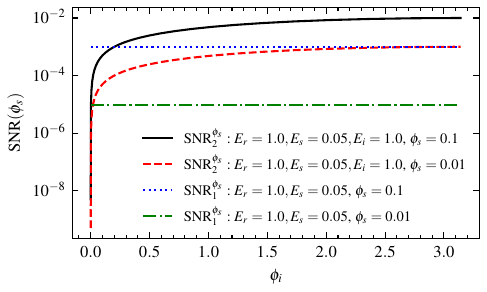} }}%
    \caption{(color online) The SNR for estimating mass (a) or estimating phase shift due to the particle (b) for different parameters.} 
\end{figure}
In Fig.~\ref{fig:snr1}, we show the $\mathrm{SNR}$ as a function of $\phi_{\mathrm{i}}$ for a signal used to analyze mass. It is clear that better performance is achieved when optimizing the second arm of the interferometer, compared to the conventional iSCAT setup. Furthermore, Fig.~\ref{fig:snr2} confirms that accurate phase estimation for small particles is possible with the second arm. In both cases, these results are consistent with our calculations of the CRB.

\bibliographystyle{apsrev4-2}
\bibliography{apssamp_emre}